# Label-Free Intraoperative Imaging of Hemodynamics using Deep Learning

YAN SHI[1], DENGHUI ZHAO[1], JINGYI YU[1], WEI NI[2], PENGCHENG LI[3], YUN GU[4,*], PENG MIAO[1,*] AND SHANBAO TONG[1]

[1]*School of biomedical engineering, Shanghai Jiao Tong University, 800 Dongchuan Road, 200240, Shanghai, China*
[2]*Department of Neurosurgery, Huashan Hospital, Fudan University, 12 Middle Wulumuqi Road, 200040, Shanghai, China*
[3]*School of biomedical engineering, Hainan University, 58 Renmin Avenue, 570228, Hainan, China*
[4]*Institute of medical robotics, Shanghai Jiao Tong University, 800 Dongchuan Road, 200240, Shanghai, China*
*\*yungu@ieee.org; pengmiao@sjtu.edu.cn*

**Abstract:** Intraoperative visualization of hemodynamics is crucial for accurate diagnosis and informed surgical decision-making. In neurosurgery, indocyanine green fluorescence imaging (ICG-FI) is the gold standard for assessing blood flow and identifying vascular structures. However, it is limited by time-consuming data acquisition, mandatory waiting periods, potential allergic reactions, and operational complexities. Label-free alternatives, such as laser speckle contrast imaging (LSCI) and white light imaging (WLI), offer real-time vascular assessment but cannot resolve arterial-venous differentiation or blood flow direction determination. To address these challenges, we present a label-free cross-modal generation framework to synthesize mean transition time (MTT) maps from LSCI and WLI. MTT maps encode local hemodynamics, enabling artery-vein differentiation and flow direction inference. Experimental validation in rat brains demonstrates that the proposed method presents clear vasculature delineation, accurate artery-vein differentiation, and reliable blood flow direction decoding, while reducing total imaging time by 95.8% compared to conventional ICG protocols. This approach offers a fast, efficient, and contrast-free solution for continuous intraoperative surgical guidance.

## 1. Introduction

Vascular imaging has become an indispensable tool across diverse surgical specialties, as precise delineation of vascular structures and accurate assessment of hemodynamics are closely associated with enhanced surgical precision and reduced perioperative complications [1-3]. In neurosurgery, conventional intraoperative vascular imaging, including intraoperative digital subtraction angiography (iDSA) [4], intraoperative magnetic resonance imaging (iMRI) [5], and Doppler ultrasonography (dUS) [6], provides valuable structural and hemodynamic information but is limited by invasiveness, prolonged acquisition time, or complex equipment requirements.

Optical imaging techniques present an appealing alternative due to their non-invasiveness, high spatial and temporal resolution, and compatibility with surgical workflows [7-9]. Among these, indocyanine green fluorescence imaging (ICG-FI) [10] has emerged as the clinical gold standard for visualizing hemodynamics through dynamic parameter mapping [11-13]. It supports arterial–venous differentiation, evaluation of microvascular perfusion, and assessment of vascular anastomoses [14, 15] and is widely used in aneurysm clipping, arteriovenous malformation (AVM) resection, and bypass verification. However, its utility is hindered by several intrinsic limitations: the administration of contrast agents introduces risks of allergic reactions and patient-specific contraindications [10, 11, 16], while logistical challenges—such as lengthy imaging protocols, mandatory waiting intervals between injections, and operational complexity—impede real-time or repeated intraoperative application [13, 17, 18]. Moreover, signal fidelity can be compromised by factors such as illumination geometry, tissue absorption and scattering, camera parameters, and dye leakage into the interstitial space, all of which can distort fluorescence sequential signals [11]. These challenges have motivated the exploration of real-time alternatives that preserve hemodynamic information without exogenous contrast agents.

Label-free optical modalities such as laser speckle contrast imaging (LSCI) and white light imaging (WLI) are promising in this context [19-22]. LSCI enables real-time, continuous mapping of blood flow by exploiting the interactions of coherent light with moving erythrocytes, providing valuable dynamic information on microvascular perfusion [23]. WLI offers high-resolution anatomical visualization of vessel morphology and surrounding tissue state, facilitating the identification of structural features critical for surgical guidance [22]. However, LSCI alone cannot differentiate arterial from venous structures or determine the directionality of blood flow, while WLI provides only static structural information and lacks intrinsic hemodynamic information.

Recent advances in deep learning offer a compelling solution by enabling cross-modal synthesis of target-modality images from readily available inputs [24], e.g., generating CT images from MRI scans [25, 26] or histologically stained images from optical microscopy data [27, 28]. Such methods facilitate the integration of structural and functional information, streamline clinical workflows, and reduce patient risk. Their feasibility arises from the intrinsic correspondence among multimodal images, which capture complementary aspects of the same anatomy and exhibit strong spatial correlations that deep networks can model. Consequently, cross-modal generation provides a promising pathway to bridge the functional gap between LSCI/WLI and ICG-FI, enabling the synthesis of hemodynamic maps from label-free data.

Among various hemodynamic indicators derived from ICG-FI, the mean transit time (MTT) map represents the temporal centroid of the fluorescence intensity curve, reflecting the average time required for blood to traverse a local vascular region [29]. Physiological differences in MTT between arteries and veins enable arterial-venous differentiation. At the same time, the spatial distribution of MTT values along vessels forms a temporal gradient that can be used to infer blood flow direction. Therefore, cross-modal generation of MTT maps using deep learning can effectively compensate for the limitations of LSCI and WLI, enabling label-free and continuous visualization of spatiotemporal hemodynamic features.

In this study, we employed a Mixed-Attention Dense UNet (MA-DenseUNet) to generate MTT maps from LSCI and WLI data. This UNet variant integrates densely connected and mixed-attention blocks to capture both local and global context. Experimental validation using multimodal imaging on the rat brain demonstrated that: (1) synthetic MTT maps generated from label-free inputs achieve robust arterial-venous differentiation and accurate flow direction decoding; and (2) the overall imaging workflow is substantially accelerated, reducing total imaging time by 95.8% compared to conventional ICG protocols. This paradigm eliminates contrast-related delays while enabling continuous, real-time surgical guidance, laying the groundwork for label-free hemodynamic mapping in neurosurgery and other vascular interventions.

## 2. Methods and Materials

### 2.1. In-vivo Data Acquisition

#### 2.1.1. Animal preparations

The animal experiment protocol was approved by the Institutional Animal Care and Use Committee (IACUC) of Shanghai Jiao Tong University (No. A2023085) on July 15, 2023. Imaging data were acquired from adult male Sprague-Dawley (SD) rats (n=23), aged 4~6 weeks and weighing 250~350 *g*. The rats were initially anesthetized with 5% isoflurane for five minutes, and anesthesia depth was monitored through tail and toe pinch reflexes. Once under deep anesthesia, each rat was secured on a stereotaxic frame to maintain head stability. Anesthesia was maintained at 2% isoflurane during surgery. After shaving the head, a central scalp incision was made to expose the calvarium. Bilateral cranial windows were created by removing as much of the parietal bones as possible. The length of the windows (anterior to posterior) followed the sagittal plane, spanning from Bregma to Lambda [30], approximately 8 *mm*. The width (medial to lateral) began 1 *mm* lateral to the sagittal suture and extended about 4*mm* outward. The sagittal suture was preserved to protect the superior sagittal sinus. Skull thinning and removal were performed using a skull drill with saline added every 15 seconds to prevent thermal damage. Finally, ICG dye (25 *mg* dissolved in 50 *mL* saline) was injected at a dosage of 2 *mL/kg* via the tail vein.

### 2.1.2. Imaging equipment and settings

All imaging was performed with a modified surgical microscope. The schematic diagram and photograph of the system are illustrated in Fig. 1(a) and (b), respectively. The system integrates a 3-CMOS camera (FS-3200T-10GE-NNC, Jai, Japan) equipped with three Sony Pregius IMX252 CMOS sensors (1/1.8 *inch*, 2048×1536 pixels) for capturing three spectral bands (visible 400~670 *nm*, NIR1 700~800 *nm*, NIR2 820~1000 *nm*). The white-light imaging outputs the color images with RGB channels. The monochromatic NIR1 sensor is used for the LSCI, while the NIR2 sensor is used for ICG-FI. Illumination was provided by a multi-modality light source, combining a white-light LED (5500 *K*, 2100 *lm*, EndoView Inc., China) and a 785 *nm* diode laser (Thorlabs, USA) via a fiber bundle. The microscope's magnification was set to 2.5× to cover the entire brain.

### 2.1.3. Data acquisition protocol

The workflow for data acquisition is illustrated in Fig. 1(c). The white light images and raw speckle images were acquired through the white light sensor (15 *ms* exposure time) and NIR1 sensor (5 *ms* exposure time), respectively. The whole recording lasted for 3 *s* at a frame rate of 60 *fps*. Auto-white-balance mode was enabled to correct the chromatic aberration in white light imaging. ICG-FI was then performed to record a video of the target cerebral vessel region, capturing the gradual fluorescence illumination process. The ICG solution was injected into the rat, and a 785 *nm* laser was used for excitation. The NIR2 sensor recorded the fluorescence image sequences with an exposure time of 15 *ms* at 60 *fps*. The camera gain was set to 16× to improve the sensitivity of fluorescence imaging.

### 2.1.4. Data Processing

Following LSCI principles, contrast images were obtained using temporal laser speckle contrast analysis (tLASCA). For each contrast image, 100 consecutive raw speckle frames were processed to compute the first- and second-order temporal statistics:

$$K = \frac{\sigma_t}{\mu_t}, \tag{1}$$

where $K$ denotes the temporal speckle contrast, $\sigma_t$ the standard deviation, and $\mu_t$ the mean intensity within the time window. The contrast value $K$ inversely correlates with blood flow velocity ($v \propto 1/K^2$). The resulting speckle contrast image is further normalized using the histogram equalization algorithm before being used as input for a deep neural network.

The MTT image is calculated pixel-wise from ICG fluorescence video:

$$MTT = \frac{\sum_t t \cdot I(t)}{\sum_t t}, \tag{2}$$

representing the temporal centroid of the intensity curve from ICG arrival to signal stabilization. The computed MTT map is then normalized via histogram equalization.

Spatial registration among the three modalities is performed to ensure alignment between input and ground truth pairs. Regions of interest (ROIs) covering the cranial window of the hemisphere are manually selected to retain only the vascular regions, excluding surrounding tissue and skull.

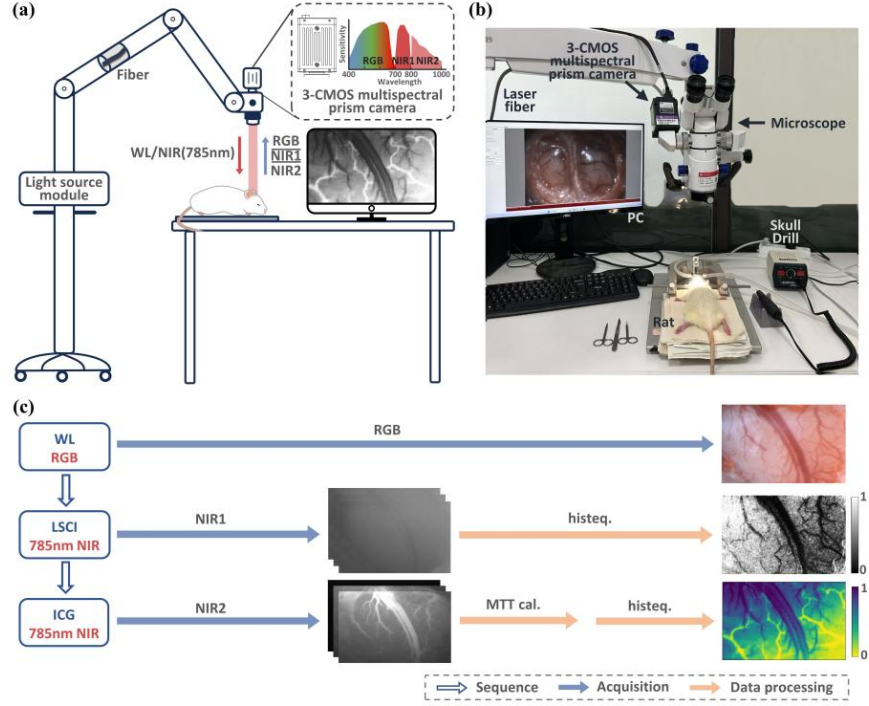

Fig. 1. Experimental setup and data workflow. (a) Schematic diagram of the intraoperative multimodal optical imaging system, integrating a 3-CMOS camera to enable white light imaging (WLI), laser speckle contrast imaging (LSCI), and ICG fluorescence imaging. (b) Photograph of the assembled system and Photograph of the assembled system and the rat brain imaging experimental scene. (c) Data acquisition and processing pipeline.

### 2.2. Cross-modal Image Generation via Deep Neural Network

#### 2.2.1. Neural Network Design

To generate MTT images from label-free modalities, we employ a UNet variant enhanced with dense connections [31] and mixed attention mechanisms (Mixed Attention Dense UNet, MA-DenseUNet, as shown in Fig. 2). The model receives a four-channel input, concatenating the white-light image and the contrast image, and outputs a single-channel MTT map of the same spatial size.

The encoder-decoder architecture with skip connections is designed to capture multi-scale features and improve feature fusion. The dense modules (Fig. 2(b)) facilitate feature reuse and help mitigate overfitting by concatenating features from preceding layers, while the mixed attention modules (Fig. 2(c)) apply channel and spatial attention sequentially to emphasize informative features adaptively. Deep supervision along the decoder further refines feature representation and accelerates convergence [32], and the final up-sampling enables computation of the vessel-weighted loss function $L_{total}$.

The loss function incorporates vessel sensitivity to guide image-to-image training of MA-DenseUNet, as follows:

$$L_{total} = \sum_{i=1}^{4} \alpha_i \cdot \left(\frac{1}{N}\sum_{j \in P}(y_j - \hat{y}_{ij})^2 + \frac{1}{M}\sum_{k \in M_v}(y_k - \hat{y}_{ik})^2\right), \quad (3)$$

where $\alpha_i$ represents the weight for each decoder layer, valued at 0.125, 0.125, 0.125, and 0.5, respectively. $N$ and $M$ represent the pixel counts for the full patch $P$ and the vasculature mask $M_v$, respectively. The pixel value $y_*$ and $\hat{y}_{i*}$ correspond to ground truth and prediction for layer $i$, with the subscripts $j$ and $k$ indexing pixels in $P$ and $M_v$, respectively. The vasculature mask is created by adaptive segmentation of the maximum-intensity fluorescence image, followed by manual refinement.

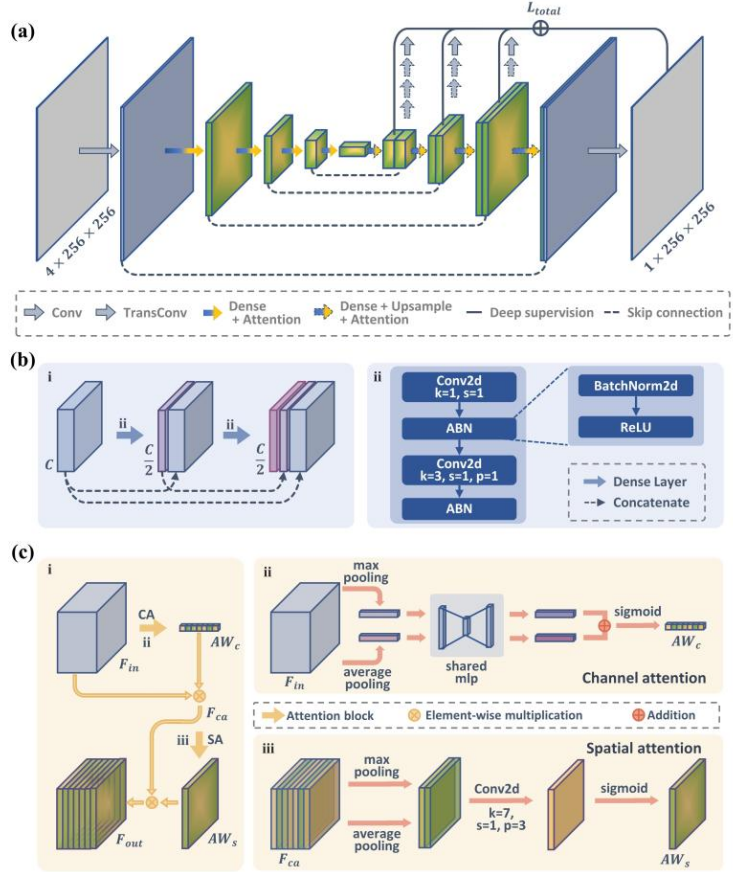

Fig. 2. Neural network design. (a) The overall structure of MA-DenseUNet incorporates dense modules and mixed attention (MA) modules in each layer. (b) Dense module structure. (c) Mix the attention module structure.

### 2.2.2. Dataset

The dataset comprises 46 samples collected from rat brain hemispheres, each containing preprocessed and paired multimodal images. Among these samples, five are excluded due to preparation issues, and ten others are designated as the Test_Poor dataset due to insufficient ICG perfusion in the ground truth MTT images, despite having qualified input images (see Fig. 3, labeled *Test_Poor1* and *Test_Poor2*). Rat randomly splits the remaining 31 samples into a training dataset (25 samples from 15 rats) and a test dataset (the Test_Normal dataset, 6 samples from 3 rats).

For training, a 10-fold cross-validation method is applied, along with data augmentation to reduce overfitting. During each training epoch, 200 patches were randomly sampled from each data set based on various transformations, including random rotation (angle $\theta_r \in [-10, 10]°$), random cropping (side length $a$=256), and random flipping (60% probability).

### 2.2.3. Implementation details

The model is implemented in PyTorch IDE and trained on an NVIDIA RTX 4070Ti GPU card with 12GB memory. The training and inference are conducted using the open-source platform *Detectron2*, developed by Meta Inc. The AdamW optimizer was employed with parameters

$\beta_1$=0.9, $\beta_2$=0.99, $\lambda$=0.01. The initial learning rate was set to 0.01 and decreased by a factor of 0.1 at first-, half-, and third-quarter epochs. We conduct 100 epochs in the training procedure. Training spans 100 epochs with a batch size of 10.

To assemble the full prediction image, we apply a Gaussian-weighted patch-combining strategy with a 2D Gaussian weight ($\sigma$=0.25) per patch, which helps maintain prediction consistency and reduces boundary artifacts. Each patch, sized 256×256 pixels, overlaps with neighboring patches by a stride of 128 along both axes.

### 2.3. Generating ICG fluorescence-like video

The generation of ICG fluorescence-like video is achieved through the utilization of a synthetic MTT image. The MTT image represents vascular perfusion time, where a smaller intensity value indicates an earlier appearance of the fluorescence signal in the ICG video sequence. Thus, the intensity values in the image can be used to convert a 2D image into a time series.

For a synthetic MTT image (intensity $I_{MTT} \in [0, 1]$), we first convert to 256-bit grayscale. The intensity scaling requires temporal alignment with the original ICG video by identifying two key timepoints: vessel appearance ($T_s$) and full perfusion ($T_e$). The frame number $N_f$ during this interval is calculated as $N_f = fr \times (T_e - T_s)$, where the frame rate $fr$=60 $fps$ in our experiment. $N_f$ is typically 60~70 frames, significantly fewer than 256 grayscale levels. Assuming the fluorescence signal in the vessels increases uniformly over time, the frame image at time $T_s + t$ corresponds to all MTT pixels satisfying $I_{MTT} \leq t \times \lfloor 256/N_f \rfloor$:

$$frame_{MTT}(T_s + t) = I_{MTT}(I_{MTT} \leq t \times \lfloor 256/N_f \rfloor). \qquad (4)$$

This base image $I_{MTT}$ can further be enhanced by replacing it with the contrast image $I_{LSI}$ and applying vessel segmentation for better visualization.

## 3. Results

### 3.1. Label-free Cross-modal Generation Enables MTT Mapping for Vascular Differentiation

Both full-reference and no-reference image quality assessments [33] are utilized to evaluate the synthetic MTT images. Full-reference assessments consist of the mean squared error (MSE) computed over the entire image and vessel-specific regions (vMSE), as well as the structural similarity index (SSIM) measure for the vessel area (vSSIM). No-reference indicators include CNR (Eq.($CNR = 20\log(|\mu_v - \mu_b|/\sigma_b)$, (5) $L_{total} = \sum_{i=1}^{4} \alpha_i \cdot \left(\frac{1}{N}\sum_{j\in P}(y_j - \hat{y}_{ij})^2 + \frac{1}{M}\sum_{k\in M_v}(y_k - \hat{y}_{ik})^2\right)$, (3) and SNR (Eq.($SNR = 20\log(\mu_v/\sigma_b)$, (6)):

$$CNR = 20\log(|\mu_v - \mu_b|/\sigma_b), \qquad (5)$$

$$SNR = 20\log(\mu_v/\sigma_b), \qquad (6)$$

where $\mu_*$ and $\sigma_*$ represent mean and standard error, with subscripts $v$ and $b$ for vascular and background areas.

The MTT image derived from ICG-FI (Eq.($MTT = \frac{\sum_t t \cdot I(t)}{\sum_t t}$,(2)) serves as the ground truth, enabling end-to-end model training from the white light and contrast images. Fig. 3 presents the generated results for two representative cases selected based on the quality of their MTT ground truth: one of high quality and one of low quality. The first four columns display the white light images, contrast images, generated MTT maps, and corresponding ground truth. In contrast, the last column highlights the regions of interest (ROIs) for detailed inspection. In the generated MTT images, arteries and veins are accurately represented, as confirmed by the vascular regions in $L_1$ error maps between the outputs and ground truth (Fig. 3, fifth column). Vessel-specific metrics further demonstrate performance on the *Test_Normal* dataset, with vMSE = 0.03 ± 0.01, vSSIM = 0.60 ± 0.04, CNR = 3.77 ± 0.48 and SNR = 7.10 ± 0.25.

In neurosurgical procedures, distinguishing arteries from veins is crucial. For label-free imaging, subtle color differences arising from hemoglobin absorption can be observed in the white light images; however, the MTT image provides a more precise and more accurate visualization (e.g., A1 and V1 in *Test_Normal1*). Another challenging task in intraoperative applications is determining branching relationships. For example, the origin of the A1 vessel within the white-dotted circles is hardly discernible in the white light image or contrast image. In contrast, both the ground truth and the synthetic MTT images reveal the branching relationships successfully.

A standard limitation of intraoperative ICG-FI is poor perfusion due to background fluorescence, leakage, and motion artifacts, which can lead to incomplete or blurred vessels (e.g., V2 in the ground truth of Fig. 3) and reduced CNR and SNR (Table 1; CNR and SNR for *Test_Poor* case are markedly lower than those for *Test_Normal* case). In contrast, cross-modal generation from label-free real-time modalities overcomes these issues. When input images are of sufficient quality, the generated MTT images present complete vessels with lower noise compared to the experimental ground-truth images. For example, the synthetic results achieve substantially higher CNR and SNR than the ground truth for *Test_Poor1* case: CNR = 6.00 vs. -0.41 dB, and SNR = 14.18 vs. 8.27 dB. These results demonstrate the robustness of label-free cross-modal MTT mapping in addressing common intraoperative challenges.

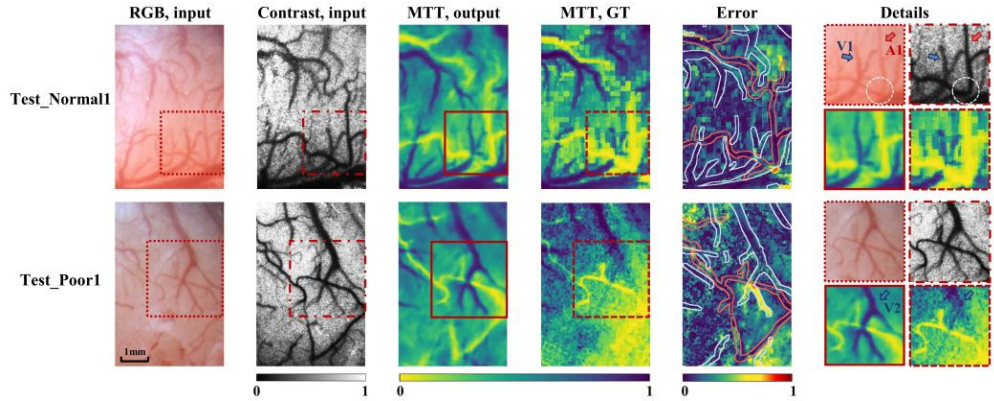

Fig. 3. Synthetic MTT images generated by MA-DenseUNet. Columns show (from left to right): (1) white light images, (2) contrast images, (3) synthetic MTT predictions, (4) ground-truth MTT, (5) L1 error maps, and (6) zoomed ROIs for comparison. Rows display one normal (top) and one poor-quality (bottom) test case. Arterial/venous boundaries in error maps are marked by red/white contours, respectively.

Table 1. Comparison Between Ground-truth and Synthetic MTT Images

| Samples | CNR (dB) | | SNR (dB) | |
|---|---|---|---|---|
| | GT | Output | GT | Output |
| **Test_Normal1** | 4.0256 | 6.0122 | 10.9609 | 13.1460 |
| **Test_Poor1** | -0.4138 | 5.9974 | 8.2694 | 14.1766 |

An ablation study is conducted to evaluate our MA-DenseUNet against three alternative models: UNet (baseline), DenseUNet with deep supervision, and MA-DenseUNet without deep supervision (*Supplementary Document S1*). Table S1 summarizes both full-reference metrics (MSE, vMSE, and vSSIM) and no-reference metrics (CNR and SNR) for each model. Fig.S1 presents the generated MTT image of the case *Test_Normal2* alongside its ground truth and corresponding attention maps, together with vessel profile comparisons and detailed peak analysis (Tables S2 and S3). Our model outperforms all alternative models, achieving the best evaluation performance with 17~33% reduction in vMSE, ~4% improvement in vSSIM and 10~54% in CNR, resulting in clearer, more continuous, and better-resolved vascular structures. Analysis of attention maps confirmed that our model effectively allocates focused attention to all vessels, which contributed to its enhanced ability to capture fine vascular details and maintain vessel continuity, particularly for weaker vessels that were poorly represented in other models. These results highlight the effectiveness of the dense and mixed-attention blocks combined with deep supervision in improving MTT map quality.

*3.1.1. Decoding Flow Direction from Synthetic MTT Image*

Determining the direction of blood flow is critical during surgery, as it assists surgeons in identifying upstream and downstream blood flows and detecting any malfunctioning vessels. For example, surgeons need to occlude the upstream vessel promptly in the event of bleeding downstream, facilitating timely hemostasis, which could be more efficient if the blood flow direction is known. In ICG-FI, the direction of blood flow is identified by directly analyzing a fluorescence video. The blood flow direction can also be determined from the $T_{1/2max}$ (also termed delay time) or MTT values [34, 35]. $T_{1/2max}$ is the time for the fluorescent brightness to rise from baseline to half of the peak intensity. In a single vessel, lower $T_{1/2max}$ or MTT values imply earlier ICG arrival, whereas higher values correspond to later arrival. Here, we find that the synthetic MTT images demonstrate comparable performance in decoding the direction of blood flow.

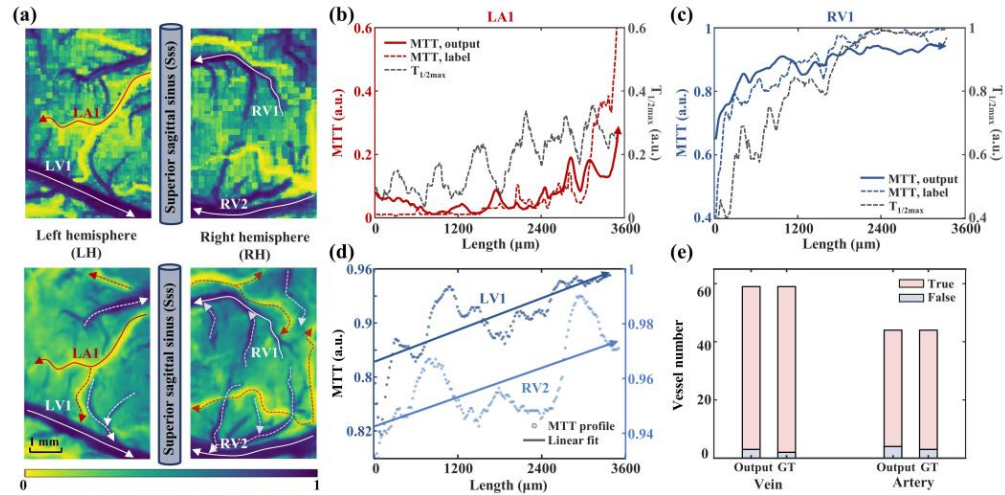

Fig. 4. Blood flow direction analysis using synthetic MTT images. (a) Synthetic MTT images (bottom) versus ground truth (top) from bilateral cerebral hemispheres, with arrows indicating flow direction. (b, c) MTT and $T_{1/2max}$ profiles along vessels LA1/RV1, showing directional flow (low→high values). $T_{1/2max}$ is another indicator of blood flow direction [35]. (d) MTT profiles of LV1/RV2 with linear fits, demonstrating drainage into the superior sagittal sinus. (e) Quantitative accuracy of flow direction detection using MTT and $T_{1/2max}$.

Fig. 4(a) shows the synthetic MTT images (bottom row) and their corresponding ground truth (top row) from both hemispheres of a rat. According to the anatomy of brain vasculature, the superior sagittal sinus (Sss), located within the superior margin of the falx cerebri, collects blood from the cerebral hemispheres. Fig. 4(b) and Fig. 4(c) show the profiles of MTT and $T_{1/2max}$ values along the vessel LA1 and RV1, denoted by arrows in both synthetic and ground-truth MTT images. According to the trends of MTT and $T_{1/2max}$ signals, we can thus infer arteries when the blood flows from the main vessel to the branches, but veins if the blood flows in the opposite direction, which complies with the circulation anatomy as well.

Additionally, the two most prominent veins in each hemisphere, namely LV1 and RV2, which are branches of the Sss, are also compared. The MTT value profiles and the corresponding linear fitting results along the two veins (indicated by arrows in Fig. 4(a)) are plotted in Fig. 4(d). These profiles illustrate blood drainage from each hemisphere into the Sss, consistent with anatomical expectations. Furthermore, the slope of the linear fitting is a robust indicator for blood flow determination. A positive slope indicates that blood flows from the start of the vessel curves toward its end, and vice versa for a negative slope. Using the $T_{1/2max}$ image as a reference, the accuracy of the slope indicator is summarized in both the synthetic and ground-truth MTT images, including a total of 59 veins and 44 arteries in the test set, as shown in Fig. 4(e). The detection of blood flow direction in synthetic MTT images achieves an accuracy of 94.92% for veins and 90.91% for arteries, respectively. The dashed lines with arrows in Fig. 4(a) indicate the blood flow directions of all remaining vessels in the two hemispheres, as determined from the synthetic MTT.

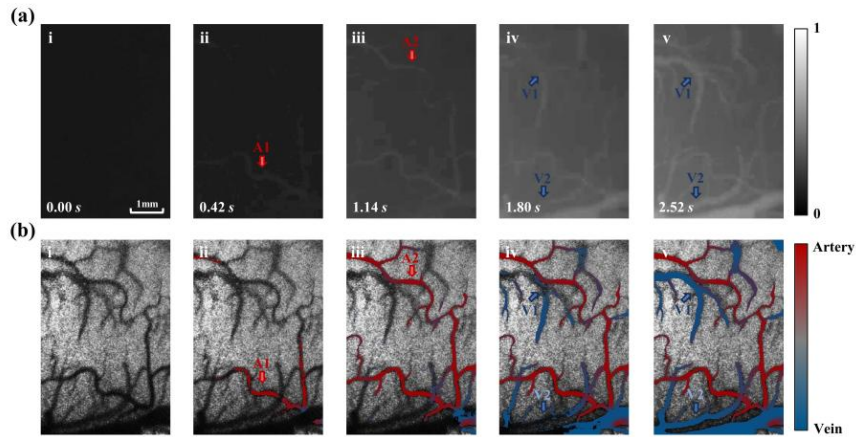

Fig. 5. Key frames comparison between ground-truth ICG video (a) and synthetic ICG fluorescence-like video superimposed on the contrast image (b). Frames (i) to (v) are displayed in chronological order.

Moreover, during an intraoperative procedure, surgeons often rely on visually observing ICG fluorescence videos to estimate blood flow velocity and direction. This practice can be both subjective and time-consuming. To streamline this process and enhance usability, we generate an ICG fluorescence-like video from synthetic MTT maps and overlay them onto the contrast images. Fig. 5 presents key frames from the synthetic ICG fluorescence-like video, superimposed on the contrast image, alongside the corresponding ground-truth ICG video (*Video S1* for details). The synthetic video captures both arterial and venous phases of hemodynamics. The artery A1 first appears in the field of view, with blood flowing into its branches (Fig. 5(a-ii) and (b-ii)). The artery A2 subsequently becomes visible at 1.14 seconds (Fig. 5(a-iii) and (b-iii)), just 0.7 seconds after A1, a time interval that is too short for human visual perception to resolve. Venous return is visualized through veins V1 and V2, with blood flowing from the branches to the trunk (Fig. 5(a-iv) and (b-iv)). Notably, V2 illuminates slightly earlier than V1, highlighting subtle differences in venous hemodynamics. In the ground-truth ICG video, arterial fluorescence fades as venous illumination progresses, potentially obscuring critical hemodynamic information. In contrast, our synthetic video preserves this information by superimposing MTT values on the contrast images, ensuring continuous visibility of arterial and venous dynamics (Fig. 5(a-v) and (b-v)).

### 3.1.2. Label-free Cross-modality Generation Accelerate Intraoperative Vascular Imaging

Cross-modality generation from real-time, label-free imaging modalities enables rapid intraoperative vascular imaging, overcoming limitations of traditional ICG fluorescence that requires 40-second acquisitions and 15-minute intervals between injections. This method captures white light images (1/50 $s$) and speckle images (~1.67 $s$) simultaneously, using each pair to generate a synthetic MTT map via a deep neural network. With time-sliding-window processing applied to the speckle data, continuous real-time MTT imaging is achieved. Compared with ICG's 16.38 $s$ per MTT computation, deep learning reduces processing to 0.70 $s$ for a 1024 × 1024 image, shortening total imaging time from 56.38 $s$ to 2.37 $s$ (a 95.8% reduction). In arteriovenous malformation (AVM) surgeries, which typically involve 2-4 imaging sessions [36], this time efficiency translates to a potential reduction of 17-50 minutes. *Video S2* demonstrates the method's real-time capability for generating MTT images in vivo.

## 4. Discussion

Accurate and efficient intraoperative visualization of hemodynamics is crucial for guiding vascular neurosurgery. Although ICG fluorescence imaging is the clinical gold standard, its reliance on exogenous contrast agents and time-consuming imaging protocols limits real-time surgical guidance. In this study, we address these challenges by developing a label-free cross-modality generation framework to synthesize MTT maps—hemodynamics parameter maps traditionally derived from ICG fluorescence imaging—from laser speckle contrast imaging and white light imaging. Our approach not only eliminates the dependency on contrast agents but also significantly improves the time efficiency, providing surgeons with detailed hemodynamic information previously accessible only through ICG imaging.

In vivo experiments were performed on rat brains using multimodal vascular imaging to validate the feasibility of our label-free cross-modality framework. The results are encouraging. First, the generated MTT images reliably differentiate arteries from veins. Second, they enable accurate decoding of blood flow direction based on their physiological significance, consistent with guidance from ground-truth MTT and $T_{1/2max}$ maps. Moreover, owing to the real-time and label-free nature of the input modalities, our approach achieved a 95.8% reduction in total imaging time compared to conventional ICG-FI. Beyond vascular neurosurgery, this framework could also benefit other procedures requiring rapid blood flow assessment, such as cardiovascular surgery [37] or organ transplantation [38]. Its label-free, real-time operation reduces OR time and avoids contrast-agent risks, and it could be adapted to other imaging modalities like hyperspectral imaging or optical coherence tomography.

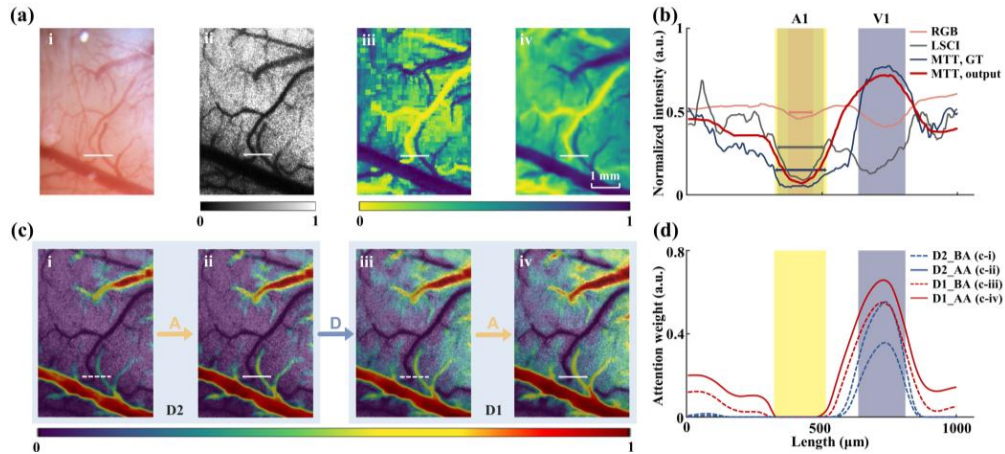

Fig. 6. Vasculature analysis across modalities with Grad-CAM visualization. (a) Comparative images: (i) white-light image, (ii) contrast image, (iii) ground-truth MTT, and (iv) synthetic MTT. (b) Normalized intensity profiles along marked vessels (gray: vein V1; yellow: artery A1). (c) Attention maps overlaid on (a-ii), showing weight distributions in the final decoder layers (A: attention module, D: dense block). (d) Quantitative attention profiles along dashed lines in (c), revealing enhanced vascular focus.

This cross-modality generation from LSCI and WLI to MTT images can be understood from both vascular structural similarity and hemodynamic feature encoding. Structurally, the major vessels exhibit consistent morphology across different modalities. As shown in Fig. 6(b), although artery A1 appears narrower in the white-light image (94.44 $\mu m$), its diameter in the contrast image (223.28 $\mu m$) is comparable to that in the ground-truth MTT image (243.55 $\mu m$), with the corresponding MTT prediction at 204.22 $\mu m$. Overall, the contrast and MTT images display similar vascular networks, highlighting their structural correspondence. Physiologically, the RGB channels in WLI capture hemoglobin-dependent spectral differences between arteries and veins, which support artery-vein discrimination during cross-modal generation. Grad-CAM (Gradient-weighted Class Activation Mapping [39]) analysis further illustrates how the network leverages these cues. Fig. 6(c) shows feature maps superimposed on the contrast images, highlighting the decoding pipeline from decoder layer 2 (D2) to layer 1 (D1). Distinct attention weights are observed in vascular regions, with arteries and veins clustering at opposite ends. Vessel peak analysis (*Supplementary Document S2*) quantitatively evaluates the attention weight profiles, as summarized in Table S4. Along the decoding pipeline, the network progressively enhances attention to both arteries and veins while sharpening the contrast between vessels and surrounding tissue. This is reflected by the increased peak-width ratio (PWR) for both A1 and V1, confirming that the network learns to prioritize vessel-specific features by jointly exploiting structural information from LSCI and spectral contrast from WLI.

Furthermore, the ability to decode blood flow direction from the predicted MTT images highlights a distinct physiological dimension. MTT values intrinsically encode the temporal dynamics of blood transit, which naturally reflect flow directionality along the vascular network. The model leverages this property during training, guided by ground-truth MTT maps as well as vascular topology such as trunk and branching relationships. As a result, it goes beyond simple pixel-level mapping to implicitly capture weak temporal constraints across pixels, thereby producing physiologically meaningful representations of blood flow direction comparable to those derived from ground-truth MTT or $T_{1/2max}$ maps. In essence, this represents weak temporal supervision, where the physiological information embedded in the ground-truth maps implicitly constrains the model. However, because this supervision is indirect, the network is not explicitly trained on temporal sequences. Future work could introduce stronger temporal supervision, e.g., by leveraging sequential fluorescence data or dynamic flow models [40], to enable more robust decoding of flow direction in complex vascular networks.

Despite these promising results, several limitations should be acknowledged. First, our current validation is limited to preclinical rat models, and clinical validation in human neurosurgery is needed to establish translational feasibility. Second, the framework currently focuses on MTT as a representative hemodynamic parameter. In contrast, integrating additional metrics, such as blood flow index (BFI) or blood volume (BV), could further enrich intraoperative assessment. Third, multimodal imaging was performed sequentially using our system, which may cause slight differences in the field of view during camera switching. Although automatic and manual registration were applied, residual misalignments could still introduce minor inaccuracies in the generated MTT maps. Fourth, while the deep learning model demonstrates great generalization within the dataset, its robustness under varying surgical conditions (e.g., bleeding, brain shift, or heterogeneous illumination) requires further investigation. Fifth, in this study, we used 100 speckle frames for contrast image calculation to improve image quality; however, 30 frames (0.5 *s* at 60 *fps*) are commonly used in clinical practice for real-time acquisition. At this lower frame rate, speckle noise is higher, which can blur vessel boundaries and cause missing information in critical vascular regions of MTT maps (Fig. S2). In future work, we aim to expand validation to multi-center clinical trials, integrate multimodal learning for richer hemodynamic representation, and optimize the imaging acquisition and inference pipeline through algorithmic improvements and hardware integration into commercial surgical microscopes, ultimately enabling real-time deployment during neurosurgical procedures.

## 5. Conclusion

In this study, we present a label-free cross-modality framework that leverages a deep neural network to generate MTT maps from LSCI and white light imaging, enabling rapid, real-time intraoperative hemodynamic assessment. Validated in rats, the method reliably differentiates arteries from veins and decodes blood flow direction, while reducing imaging time by over 95% compared to conventional ICG-FI. This approach provides a safe, efficient alternative for vascular neurosurgery and demonstrates potential for broader clinical applications in surgeries requiring fast, non-invasive, and contrast-free evaluation of blood flow.

**Funding.** This study is supported by the Fundamental Research Funds for the Central Universities (YG2023QNA27).

**Acknowledgment.** The Institutional Animal Care and Use Committee (IACUC) of Shanghai Jiao Tong University (No.A2023085) approved the animal experiment protocol.

**Data availability.** The data used in this study are available upon request, with four examples from Fig. 4 accessible at https://github.com/MPAPS2019/GenerativeMTT. For the full dataset, please contact the corresponding authors. The code, including training and inference instructions, model checkpoints, and example data, is publicly available in the same repository.

**Supplemental document.** See *Supplement 1* for supporting content.